\documentclass[11pt, twoside]{article}
\usepackage{asp2006}
\usepackage{psfig}
\usepackage{epsf}
\usepackage{graphicx}
\begin{document}
\title{Measuring Meridional Circulation in the Sun}

\author{Thomas L. Duvall, Jr.}
\affil{Solar Physics Laboratory, NASA/Goddard Space Flight Center, Greenbelt, MD 20771, USA}
\author{Shravan M. Hanasoge}
\affil{W. W. Hansen Experimental Physics Laboratory, Stanford University, Stanford, CA 94305}

\begin{abstract}
Measuring the depth variation of the meridional flows is important for understanding the solar cycle, at least according to a number of dynamo models. While attempting to extend the early observations of \citet{giles_thesis} of time-distance measurements of flow, we have stumbled upon some systematic errors that can affect these measurements: 1) the additional distance traveled by radiation coming from points away from disk center causes an apparent `shrinking' Sun, that is an apparent flow towards the disk center and 2) in measurements away from the central longitude, the rotation signal can leak into meridional flow signals. Attempts to understand and overcome these systematic problems will be presented. Forward models based on ray theory have been applied in order to test the sensitivity of travel times to various models.
\end{abstract}


\section{INTRODUCTION}\label{intro}
Meridional flow is thought to play a crucial role in determining properties of the solar dynamo, in controlling the formation and dissipation of magnetically active regions,
at least within the schema of the flux transport dynamo \citep[e.g.,][]{dynamo_review0, dynamo_review1,dikpati06}. Over the past decade, there have numerous efforts
and accomplishments with regards modeling and predicting solar cycles; among these theories, the flux transport description occupies the center stage at present. The
success of the model of \citet{dikpati06} in matching the past several cycles and the anticipation generated in the community in terms of its prediction of the
2008-  cycle was certainly exciting. In this model, the meridional flow executes the important task of transporting magnetic flux poleward from the equator, preparing
the dynamo for the onset of the next cycle. Our best observational constraints on the speed and scale of meridional flow in the Sun \citep[e.g.][]{giles97, giles_thesis} were
made possible through a combination of high quality Michelson Doppler Imager \citep[MDI;] []{scherrer95} observations and the wave statistical measurement technique
of time-distance helioseismology \citep{duvall}. Here we describe an attempt to isolate meridional flow signals through analyses of data collected by the now-decade long 
MDI medium-$l$ observational program. A number of serious systematical issues are encountered; the resolution of some issues has been possible, at the cost, it seems,
of discovering others. We discuss the nature of these problems and some forward theoretical methods applied to model the flow.

\section{COMMON MIDPOINT AND QUADRANT METHODS}\label{tdmethods}
\begin{figure}[!ht]
\begin{centering}
\vspace{-0.5cm}
\includegraphics[scale=0.35]{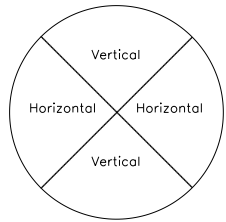}
\vspace{-1cm}
\caption{Quadrant geometry used to produce time differences. Diametrically opposite points are correlated and differenced, with the resultant termed ``east-west" or ``north-south"
based on whether the points were in the slices ``horizontal" or ``vertical". Mean travel times (common mid-point method) are computed by summing the mean correlations of all
diametrically opposite points.  \label{quadrant}}
\end{centering}
\end{figure}

The oscillations are analyzed via techniques of time-distance helioseismology \citep{duvall,duvall2003}, and in particular, mean travel times are estimated using the common-midpoint 
method of deep focusing and difference travel times with the quadrant technique. Figure~\ref{quadrant} shows an example of the quadrant geometry. Essentially points on 
opposite ends of an annulus are correlated, the directionally unbiased correlation contributing to the mean times and the differenced correlations to the difference times. 

A Gabor wavelet is fitted to these averaged correlations to obtain the phase travel times. This time is associated with the center point of the annulus; the process is repeated for all other
points. In this situation we have shown only 4 quadrants; however the analysis can be altered so that the annulus is divided into 12 slices. The 12 slice geometry appears to show
reduced inconsistency in comparison to its 4 quadrant counterpart. We discuss this issue further in $\S$\ref{errlat}.

\subsection{INSTRUMENT FLIPPING}
For a few years now, the MDI instrument onboard the Solar and Heliospheric Observatory (SOHO) has been flipped four times a year, resulting in alternating inverted views of the Sun. Of course, while this is
accounted for in the data, certain anomalies arise that are quite puzzling. We encountered large differences in the travel times in the hemispheres and most surprisingly,
around the equator, which by all considerations is viewed by the same part of the CCD camera. The effects are reduced to some extent by applying specific image distortion
corrections (Schou, J. 2008, private communication) but still persist at a level greater than the meridional flow signal strength. 

\section{FINITUDE OF THE SPEED OF LIGHT}\label{finitude}
\begin{figure}[!ht]
\begin{centering}
\plottwo{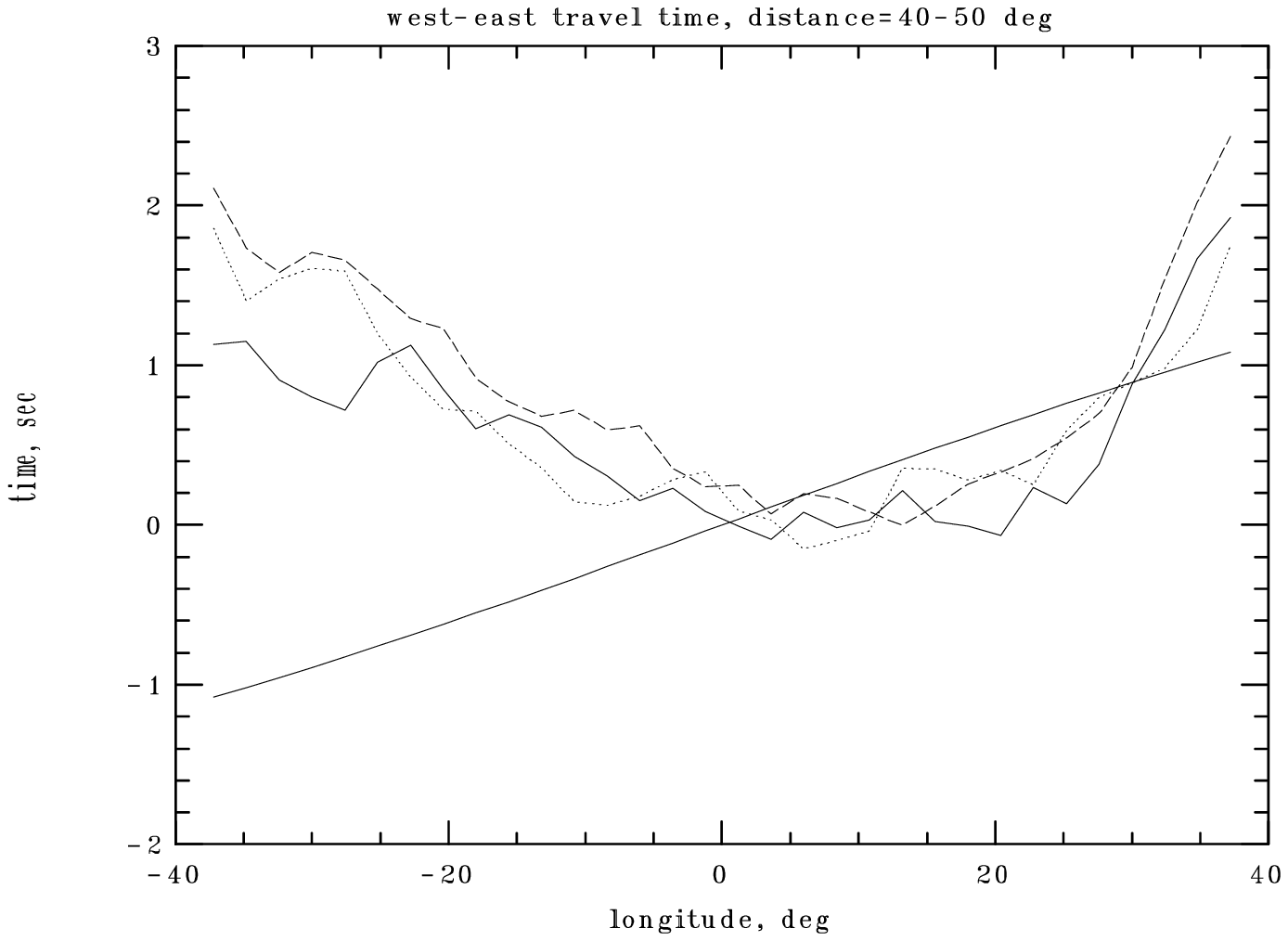}{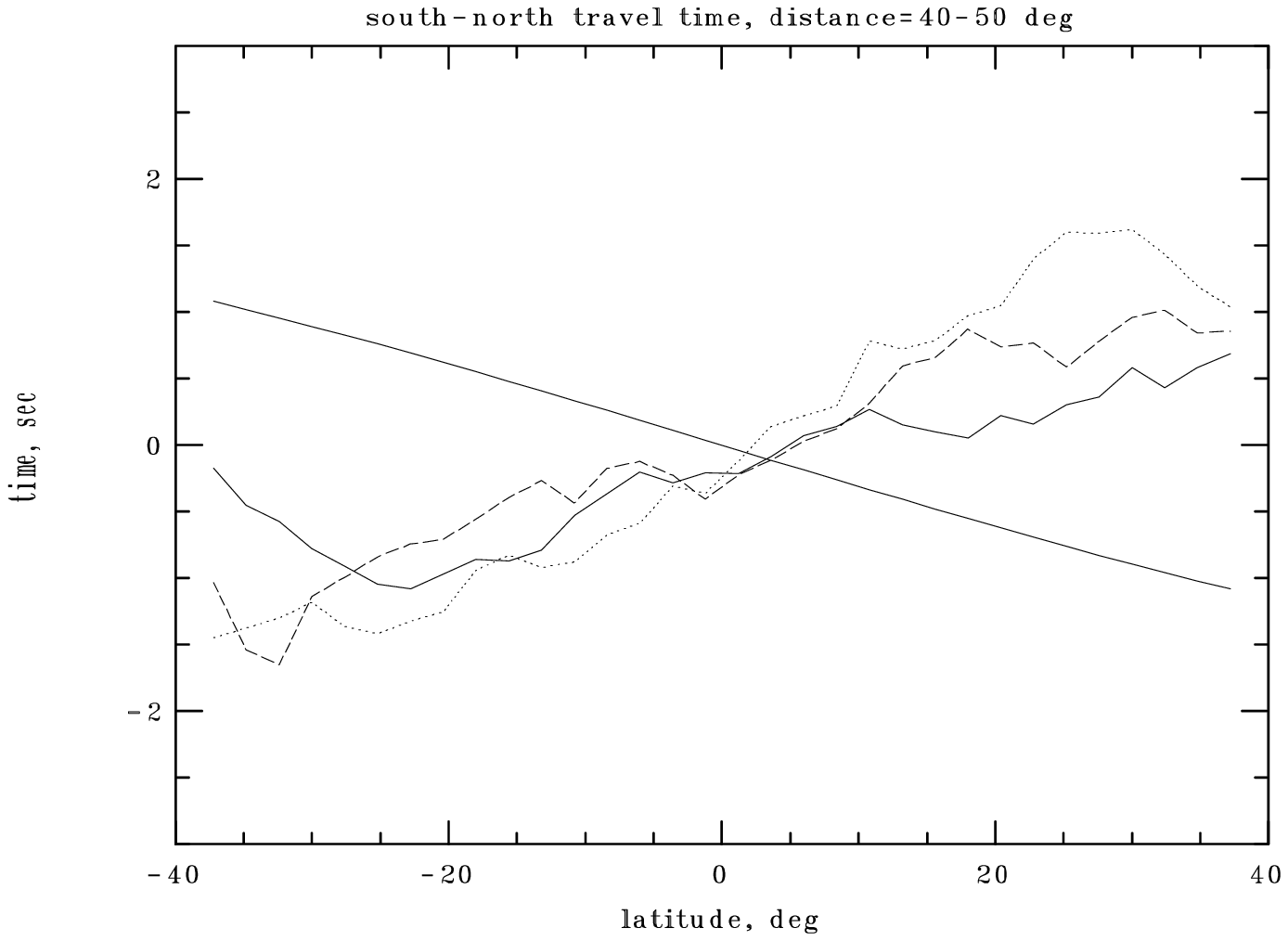}
\caption{The three curves are for two-year intervals 1996.5-1998.5 (solid), 
1999-2000 (dashed), and 2001-2002 (dotted).  The straight line is the
travel time signal expected from the light travel delay caused by the larger 
distance traveled for points closer to the edge of the solar disk.  The images
for this analysis were tracked at a solar rotation rate which removes a large
constant travel time difference. The lack of agreement between the light travel
time curve and the measured residuals remains a puzzle. On the left are the
east-west differences plotted along the equator. On the right are the north-south differences
plotted along the central meridian. The separation distance is 40-50$^\circ$.\label{lightspeed}}
\end{centering}
\end{figure}

For a far-away observer, signals at the edge of the solar disk in comparison to the disk center are delayed by not insignificant amounts due to the finite time taken by light to travel
the extra distance. The operative relation is: $R_\odot/c_{light} = 2.32$ seconds. For a variety of purposes, this does not present a hurdle; in global mode analysis of rotation for example, the error is only at 
higher order since at the zeroth, global modes propagate over the entire disk, thereby averaging out the delay. Similarly when computing mean travel times, these delays effectively 
cancel. The error starts to become quite significant when one attempts to precisely estimate time differences for waves that propagate significant distances. Because signal 
delays are present both in the north-south and the east-west directions, they mimic the signatures of an inflow towards the disk center as observed by the satellite. 
The light time delays are exacerbated with (1) increasing
heliocentric angle from the center for a given wave propagation distance, and (2) growing distance between the time-distance antennae (i.e., the wave propagation distance) for a 
given non-disk-center position. In Figure~\ref{lightspeed} (left panel) we display the east-west time difference signals averaged over a small range of latitudes with longitude. Unfortunately, for 
unknown reasons, the delays estimated from the observations are not in agreement with expectations. It appears that the residual errors are not anti-symmetric at all, and in fact seem
to be largely symmetric. These perplexing results could perhaps be explained through theories of center-to-limb related radiative heat transfer effects; regardless, we are at a loss to 
decipher the meaning of the residuals in the observed shifts.
   
 If one were to take MDI observations and perform the time-distance analyses as described in $\S$\ref{tdmethods} without paying attention to these time delays, the resultant north-south times would
 resemble the curves in Figure~\ref{lightspeed} (right panel). Time shifts of this magnitude for waves propagating distances of 40-50$^\circ$ imply a circulation of order 1 km/s or more at the base of
 the convection zone and a penetrative flow well into the radiative interior, at complete odds with our current theoretical picture of the Sun. Initial efforts were expended on attempts to
 devise a travel time correction based on the premise that the rotation rate remains constant with longitude (see $\S$\ref{theory}). However, the correction itself seemed to be vary from year to year, implying that
 under it all lay some non-trivial systematical issue that was perhaps still influencing the time signals. Furthermore, the magnitude of the correction well exceeded that of the meridional
 flow signal strength, thereby increasing the uncertainty of what we were obtaining.



\section{THEORETICAL MODELS}\label{theory}
\begin{figure}[!ht]
\begin{centering}
\includegraphics[scale=0.5]{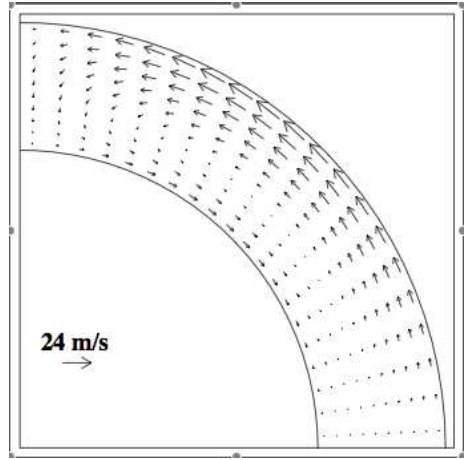}
\caption{A sample ``{\it nice}" meridional flow profile \citep{giles_thesis}.\label{flowprofile}}
\end{centering}
\end{figure}

\begin{figure}[!ht]
\begin{centering}
\plotone{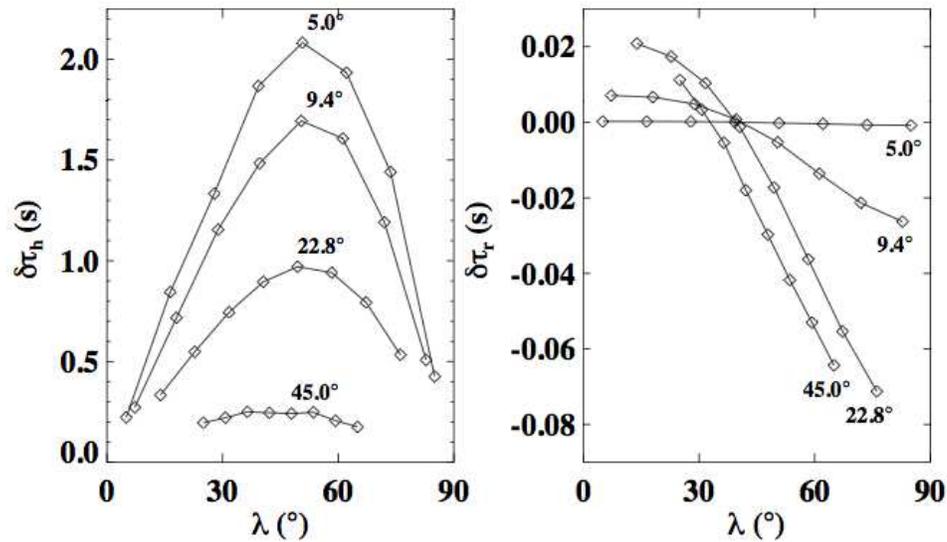}
\caption{Ray theoretic time shifts predicted for the profile of Figure~\ref{flowprofile} \citep{giles_thesis}. The left panel shows contributions to the time differences from horizontal components of
the meridional circulation and the right panel from the radial flow. For rays that travel to the bottom of the convection zone, the maximum time shift is a mere 0.2 seconds, not of the order of 1 second as seen in the uncorrected data. \label{timesprofile}}
\end{centering}
\end{figure}

The ray theoretic techniques of calculating time shifts, e.g. \citet{giles_thesis}, are relevant
in this situation because of the large spatial scales associated with meridional circulation (large in comparison to the acoustic wavelengths). The profiles of meridional flow that decay gently over
most of the convection zone and persist at the bottom with magnitudes of the order of 2 m/s or so (Figure~\ref{flowprofile}) are only possible if the north-south time shifts (Figure~\ref{timesprofile})
decay rapidly with wave propagation distance. For the time shifts shown in Figure~\ref{lightspeed} (right), the circulation velocity would have to be nearly 1 km/s at the base of the 
convection zone, and penetrate well into the radiative interior. The reason for this follows from the relation governing the time shifts:
\begin{equation}
\delta\tau = \int_{\Gamma} -\frac{{\bf u}\cdot{\hat{\bf n}}}{c^2} ds,
\end{equation}
where $\delta\tau$ is the time difference, $\Gamma$ is the unperturbed ray path, ${\bf u}$ is the imposed velocity, ${\hat{\bf n}}$ is the tangent vector to the ray, $c$ the sound speed and $ds$ the
displacement along the ray path. Because of the $c^{-2}$ term, the sensitivity of a ray to a unit flow is significantly lower at depth than near the surface. Rays that travel large distances propagate nearly
vertically close to the surface and the almost horizontal meridional flows register only weakly on these rays. At depth, not only is the
return flow (supposedly) weak but so is the sensitivity of the rays. Consequently, to sustain constant time shifts with increasing distance, the meridional flow must increase as a polynomial power of
$c$.  

\begin{figure}[!ht]
\begin{centering}
\plotone{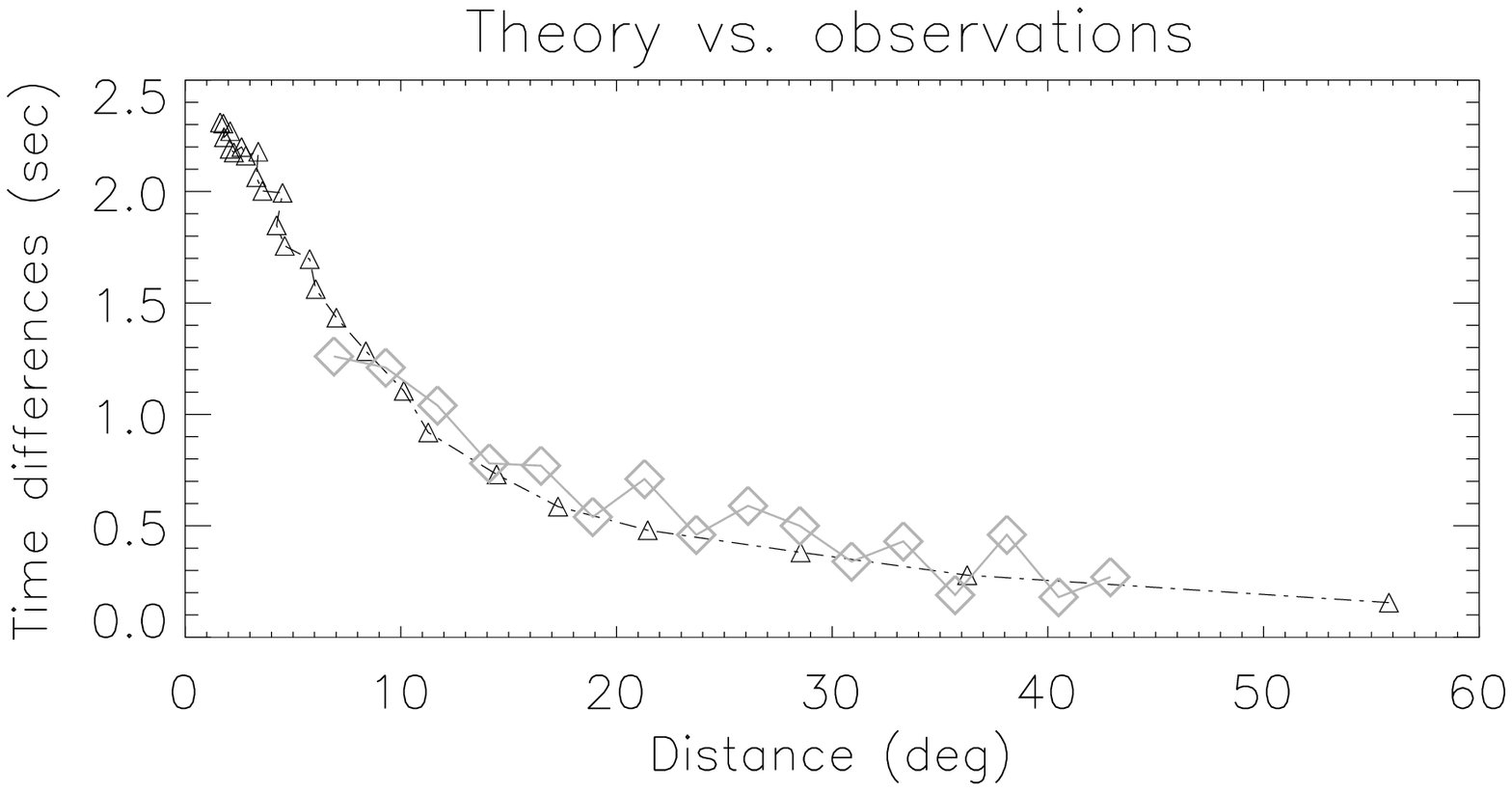}
\caption{A forward ray theoretic calculation to match observations. The diamond shaped symbols represent the longitudinally averaged, ``corrected" maximum north-south time differences over all 
latitudes, derived from observations. The triangles show the best fit theoretical times; numerous circulation models were studied before arriving at this version. The travel-time correction was designed
 on the premise that observational east-west time shifts  associated with rotation be constant with longitude ($\S$\ref{finitude}). \label{timesprofile2}}
\end{centering}
\end{figure}

 In Figure~\ref{timesprofile2}, we show results from the latest analyses of the observations. It is seen that the "corrected" observationally derived time differences are matched quite well by theory. The 
 meridional circulation profile used in these calculations contains two cells in latitude, with a dependence on both $\sin(2\theta)$ and $\sin(4\theta)$, where $\theta$ is latitude. The higher order 
 $\sin(4\theta)$ cell closes at a depth of $0.9 R_\odot$, while the return depth of the $\sin(2\theta)$ cell is more difficult to constrain and could very well be of the form seen in Figure~\ref{flowprofile}. 
 
 On another note, it may be argued that ray theory is not entirely trustworthy due to the prevalence of finite wavelength effects. To address this problem we study wave-flow interactions using 
 a wave mechanical description of the field. Having arrived at a point of reasonable agreement between ray theory and a given set of observations, we will computationally study the interactions of the 
 wave field with this flow profile using the techniques of \citet{hanasoge1}.

\section{OTHER ANOMALOUS GRADIENTS}\label{errlat}
\begin{figure}[!ht]
\begin{centering}
\includegraphics[scale=0.65]{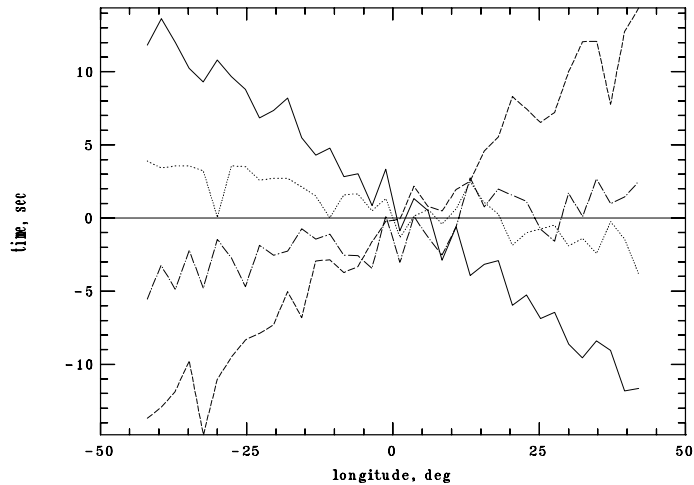}
\caption{North-south travel times for the 4-sector (quadrant) geometry versus longitude 
compared with the north-south times for a 12-sector geometry.
For the latitude range 20-32$^\circ$ north, the solid curve is for the quadrant geometry
while the dotted curve is for the 12-sector geometry.  For 20-32$^\circ$ south,
the dashed curve is for the quadrant geometry while the dot-dashed curve is
for the 12-sector geometry. The large gradient signals away from the equator are some sort of artifact.
The distance range is 12.9-13.5$^\circ$ for the quadrant geometry while the 
distances were averaged over the larger range 11.7-14.7$^\circ$.  The slopes for the
quadrant geometry are larger by a factor of 4.4.\label{lat_nasty4_12}}
\end{centering}
\vspace{-0.4cm}
\end{figure}

Because we do not expect meridional flows to vary with longitude, we typically average the north-south time shifts over longitude, the gain being that of a
surplus in the signal-to-noise ratio. However, a more careful look at the quantities that are being averaged brings out a rather sinister systematical issue, shown in Figure~\ref{lat_nasty4_12}. 
The north-south time shifts averaged over a number of latitudes close to the equator behaves quite well, showing no drift or gradient with longitude. However,
the same procedure when repeated at higher latitudes, produces large time gradients with longitude, of the order of $\pm 10$ seconds. Note that opposite hemispheres
show gradients of reversed signs. The origin of this gradient is entirely unknown at this point, except that it is presumably related to the rotation signal. We conclude this
from comparisons of north-south travel times obtained using the 12 and 4-sector geometries; the former shows much less of a gradient than the latter, as can be seen
in Figure~\ref{lat_nasty4_12}. While an average over longitude should effectively kill the gradient, small errors arising from this unknown systematic may act to reduce the 
certainty of any results derived hereof.


\section{CONCLUSIONS}
The marvelous repository of data collected since the deployment of the SOHO satellite in 1996 has certainly galvanized the solar and astrophysics community. However, with the analyses presented here, it appears
we have pushed the data to its very limit and the metaphorical cracks on the wall have started to show. The inability of current methodology and observations to constrain deep meridional flows may
appear somewhat disheartening. Progress in this regard will require the application of considerable efforts towards the resolution of this plethora of systematical issues. However, success in this endeavour 
will be rewarded by the ability to analyze a decade's worth MDI observations.

 
\bibliographystyle{asp}
\bibliography{merid}
\end{document}